\documentclass{article}
\usepackage{preprint}

\usepackage{titling}
\usepackage{lipsum}
\usepackage{comment}
\usepackage{soul}

\usepackage{authblk}

\usepackage[backend=bibtex,style=numeric-comp, sorting=none]{biblatex}
\usepackage[autostyle=true]{csquotes}
\usepackage{amsmath}
\usepackage{amsfonts}
\usepackage{mathtools}
\usepackage{bm}
\usepackage{authblk}
\usepackage{booktabs}
\usepackage{silence} 
\usepackage{subcaption}

\usepackage[hidelinks,hypertexnames=false]{hyperref}
\usepackage[all]{hypcap}
\usepackage[utf8]{inputenc}
\usepackage[T1]{fontenc}
\usepackage{caption}
\usepackage{color, colortbl}
\usepackage{microtype}
\usepackage{cleveref}

\usepackage{dsfont}
\usepackage{braket}

\usepackage{titlesec}
\titlespacing\section{0pt}{12pt plus 3pt minus 3pt}{1pt plus 1pt minus 1pt}
\titlespacing\subsection{0pt}{10pt plus 3pt minus 3pt}{1pt plus 1pt minus 1pt}
\titlespacing\subsubsection{0pt}{8pt plus 3pt minus 3pt}{1pt plus 1pt minus 1pt}

\titleformat{\section}[block]{\centering\Large\bfseries\filcenter}{}{1em}{}
\titleformat{\subsection}[hang]{\bfseries}{}{1em}{}
\titleformat{\subsubsection}[hang]{\bfseries}{}{1em}{}

\pretitle{\begin{center}\huge\bfseries}
\posttitle{\par\end{center}\vskip 0.5em}
\preauthor{\begin{center}\large}
\postauthor{\end{center}}
\predate{\par\large\centering(Dated: }
\postdate{)\par}

\definecolor{fhggray}{rgb}{0.8,0.87,0.9}

\title{\Large Unsupervised Quantum Anomaly Detection on Noisy Quantum Processors}

\author[1]{Daniel Pranji\'{c}}
\author[1]{Florian Kn\"{a}ble}
\author[1]{Philipp Kunst}
\author[1]{Damian Kutzias}
\author[1]{Dennis Klau}
\author[1]{Christian Tutschku}
\author[2]{\\Lars Simon}
\author[2]{Micha Kraus}
\author[2]{Ali Abedi}

\affil[1]{\textit{Fraunhofer IAO,  Nobelstraße 12, 70569 Stuttgart, Germany}}
\affil[2]{Bundesdruckerei GmbH, Kommandantenstraße 18, 10969 Berlin, Germany}

\date{\today}
\begin{document}

\twocolumn[
	\begin{@twocolumnfalse}

\maketitle
\thispagestyle{empty}

\begin{abstract}

\noindent Whether in fundamental physics, cybersecurity or finance, the detection of anomalies with machine learning techniques is a highly relevant and active field of research, as it potentially accelerates the discovery of novel physics or criminal activities. 
We provide a systematic analysis of the generalization properties of the One-Class Support Vector Machine (OCSVM) algorithm, using projected quantum kernels for a realistic dataset of the latter application. These results were both theoretically simulated and experimentally validated on trapped-ion and superconducting quantum processors, by leveraging partial state tomography to obtain precise approximations of the quantum states that are used to estimate the quantum kernels. Moreover, we analyzed both platforms respective hardware-efficient feature maps over a wide range of anomaly ratios and showed that for our financial dataset in all anomaly regimes, the quantum-enhanced OCSVMs lead to better generalization properties compared to the purely classical approach. As such our work bridges the gap between theory and practice in the noisy intermediate scale quantum (NISQ) era and paves the path towards useful quantum applications. \\

\end{abstract}
\vspace{0.35cm}
	\end{@twocolumnfalse}]

\section*{I. INTRODUCTION}
Anomalies are crucial in several fields of research and applications. One well-known example in physics are deviations from the standard model (SM) that indicate new physics.
Since a lot of data is being produced in particle collisions, but the interesting events are rare, most of it is dropped immediately. Hence, successfully filtering out the data by potential anomalies is a key ingredient to unveil new physics beyond the SM. Traditional approaches for anomaly detection often rely on computationally expensive simulations or expert knowledge for rule-based systems. Lately, it has been shown that novel methods like Quantum Machine Learning (QML) offer a promising avenue for addressing this challenge, potentially surpassing classical approaches in both efficiency and accuracy \cite{liu2021, farhi2014quantum, biamonte2017quantum, harrow2009quantum}.  \\
Currently available quantum computers are noisy and small. Despite these strong limitations it is possible to run so-called hybrid quantum algorithms, that combine the power of classical and quantum computing. One such example from QML, that can be implemented with current hardware, is the One-Class (Quantum) Support Vector Machine (OC(Q)SVM) \cite{steinwart, schölkopf, ocsvm_2001, ocsvm_amer2013enhancing, ocsvm_hejazi2013one, ocsvm_yin2014fault}, that leverages quantum kernels \cite{QSVMrebentrost, QSVMcoherent, kyriienko2022unsupervised, huang}. Quantum kernels implicitly map data into higher-dimensional feature spaces, where non-linearly seperable classification tasks may become linearly seperable by some hyperplane, determined by the Support Vector Machine (SVM) algorithm. Although quantum Hilbert feature spaces have the benefit of exponentially growing dimensions with respect to the qubit number, the models trained therein may become too expressive, which might lead to overfitting. This phenomenon is known as the so-called curse of dimensionality. For this reason, reducing the dimension of the full quantum feature space by projecting back into classical space is a frequently used technique. The resulting projected quantum kernels \cite{huang, QSVMconcentration, proj_kernel_suzuki2023effect} require partial state tomography. Up to date, many recent works have addressed the theoretical properties of quantum kernel methods \cite{huang, QSVMconcentration, proj_kernel_suzuki2023effect, kernel_gil2024expressivity, kernel_schuld2021supervised, kernel_jerbi2023quantum, kernel_shaydulin2022importance, kernel_schuld2019quantum, kernel_glick2024covariant, kernel_shirai2024quantum, kernel_gentinetta2024complexity, kernel_rebentrost2014quantum} and their practical properties, i.e. dependence on hyperparameters \cite{kernel_egginger2024hyperparameter, kernel_canatar2022bandwidth}, benchmarks \cite{kernel_schnabel2024quantum, kernel_alvarez2024benchmarking, kernel_bowles2024better}, applications \cite{kernel_wu2023quantum, kernel_ruskanda2023quantum, kernel_kumar2024brain, kernel_heredge2021quantum, kernel_stuhler2024evaluating, kernel_wu2021application, kernel_wu2022application, kernel_wozniak2023quantum} and physical implementations \cite{experiment_peters2021machine, experiment_hubregtsen2022training, experiment_yang2019support, experiment_haug2023quantum, experiment_simoes2023experimental}. 

In this work we contribute to this active field of research by addressing the following crucial open questions, which are mandatory for the practical application of quantum kernel methods:

\begin{enumerate}
    \item For which kind of data sets (e.g. number of samples, classical correlations \& dimensionality) do QML models provide better generalization properties in comparison to their classical counterparts? 
    \item Does such a generalization enhancement persist in the presence of hardware noise and finite sampling effects?
    \item With what precision can classical data be encoded in and loaded from current state-of-the-art quantum computers and at what resource cost in terms of sample complexity? 
\end{enumerate}

Besides those model and data-centric questions, we address one more methodological one, crucial for the performant implementation of our QML models:

\begin{enumerate}
\setcounter{enumi}{3}
    \item How precise are state-of-the art state tomography methods on real quantum hardware and what are realistic limitations for different settings (e.g. sample complexity or basis choice)?
\end{enumerate}

Let us emphasize that the last question is not only relevant for QML, but any quantum algorithm that requires the readout of a complex state at the end of a circuit, e.g., the solution to a linear system encoded in a state $\ket{x}$, is also affected. To showcase the practical impact of our paper, we address those questions in a concrete use case setting, based on financial fraud detection. In more detail, we leverage the Sparkov toolkit \cite{Shenoy.2020} to generate a simulated, time-series dataset encompassing both normal and fraudulent transaction activities. To assess the precision of our method and gain insights into the impact of noise and imperfect calibrations on the algorithm's performance, we perform partial state tomography \cite{huang_shadows, optimizationshadows}. This technique allows us to characterize the state of the quantum system and compare it to the ideal, noise-free simulation result. Furthermore, we implement the full OCQSVM pipeline and compare the performance of trained models on both trapped-ion and superconducting quantum processing units (QPUs) with noise-free simulations. This approach enables us to control the characteristics of the data and evaluate the real-world performance of our quantum machine learning algorithm.

The remainder of the paper is structured as follows: In Sec. II.A we introduce the classical OCSVM algorithm. In Sec. II.B we define the quantum radial basis function (qrbf) kernel, which is calculated from the one-qubit reduced density matrices (1Q-RDMs), obtained by means of partial state tomography (see Sec. II.C). In Sec. III.A we present our simulation results from different OCQSVMs based on hardware efficient feature maps and compare those to the classical reference model (rbf-kernel OCSVM). We further analyze the precision of partial state tomography on superconducting QPUs in Sec. III.B. We benchmark the OCQSVM models with the feature maps from Sec. III.A on QPUs by AQT, IonQ and IBM in Sec. III.C. To provide a clear and coherent message, we shifted some details of our analysis to the App. and refer to them whenever needed.


\section*{II. METHODS}

\subsection*{A. One-Class Support Vector Machines}

OCSVMs \cite{steinwart,schölkopf, ocsvm_2001} offer a powerful tool for unsupervised anomaly detection, where the model learns from unlabeled data points to identify anomalies deviating significantly from the normal patterns. Unlike standard SVMs that require labeled data for both normal and anomalous classes, OCSVMs learn a decision boundary, enclosing the normal data in a high-dimensional feature space. Points outside this boundary are flagged as anomalies.

Given training vectors $x_i \in \mathbb{R}^n, i = 1, \dots, l$ without any class information, the primal problem of OCSVMs is defined as the following constrained optimization problem

\begin{equation}\label{eq:ocsvm_prime}
    \begin{split}
    \min\limits_{w \in \mathbb{R}^n, \, \mathbf{\xi} \in \mathbb{R}^l, \, \varrho \in \mathbb{R}} \quad &\frac{1}{2} \left|\left| w \right|\right|^2 - \varrho + \frac{1}{\nu l} \sum\limits_{i=1}^l \, \xi_i  \\
    \text{subject to} \quad &w^T \phi(x_i) \ge \varrho - \xi_i \, , \\
    & \xi_i \ge 0 \, , i=1,\dots l \, ,
    \end{split}
\end{equation}
where $\phi(x_i)$ maps $x_i$ into a higher-dimensional feature space and $\xi_i$ are the slack variables for allowing soft margins, i.e. outliers outside of the classification boundary. The parameter $\nu \in (0, 1]$ was proven to be an upper bound on the fraction of training errors and a lower bound of the fraction of support vectors \cite{nusvm2000}.

Because of the possible high dimensionality of $w$, in practice, the (rescaled) dual problem \cite{chang2011libsvm} is solved instead. To obtain this, the generalized Lagrangian $\mathcal{L}$ corresponding to Eq.~\eqref{eq:ocsvm_prime} is considered

\begin{equation}
    \begin{split}
    \mathcal{L}(w, \xi, \rho) = &\frac{1}{2} \left|\left| w \right|\right|^2 - \varrho + \frac{1}{\nu l} \sum\limits_{i=1}^l \, \xi_i \\
    &+ \sum\limits_{i=1}^{l} \, \alpha_i \left( \varrho - \xi_i - w^T \phi(x_i) \right) - \sum\limits_{i=1}^{l} \, \beta_i \xi_i \, , \label{eq:lagrangian}
    \end{split}
\end{equation}
where $\alpha_i, \beta_i \ge 0$ are Lagrange multipliers. Next, the $\mathcal{L}$ with respect to the primal variables is minimized by setting the following derivatives of $\mathcal{L}$ to zero

\begin{align}
    \nabla_w \mathcal{L}(w, \xi, \varrho) &= w - \sum\limits_{i=1}^{l} \, \alpha_i \phi(x_i) = 0 \, , \label{eq:def_w} \\ 
    \nabla_\xi \mathcal{L}(w, \xi, \varrho) &= \sum\limits_{i=1}^{l} \, \frac{1}{\nu l} - \alpha_i - \beta_i = 0 \, , \label{eq:def_beta} \\
    \nabla_\varrho \mathcal{L}(w, \xi, \varrho) &= \sum\limits_{i=1}^{l} \, \alpha_i - 1 = 0\, \label{eq:def_alpha}.
\end{align}

Eqs.~\eqref{eq:def_w}, \eqref{eq:def_beta} and \eqref{eq:def_alpha} are used to simplify the Lagrangian $\mathcal{L}$ in Eq.~\eqref{eq:lagrangian} to

\begin{equation}
    \mathcal{L}(w, \xi, \varrho) = \frac{1}{2} \sum\limits_{i=1}^{l} \, \alpha_i \alpha_j \, \phi(x_i)^T \phi(x_j) \, . \label{eq:lagrangian_2}
\end{equation}

With Eq.~\eqref{eq:lagrangian_2} the dual problem of Eq.~\eqref{eq:ocsvm_prime} can be rewritten as

\begin{equation}\label{eq:ocsvm_dual}
    \begin{split}
    \min\limits_{\alpha \in \mathbb{R}^l} \quad & \frac{1}{2} \alpha^T K \alpha \\
    \text{subject to} \quad &0 \le \alpha_i \le 1 \, , i = 1, \dots l \, , \\
    &e^T \alpha = \nu l \, ,
    \end{split}
\end{equation}
where $e\equiv(1, \dots, 1)^T$ and $K_{ij}\equiv K(x_i, x_j) \equiv \phi(x_i)^T \phi(x_j)$. A frequent choice for the kernel $K(x_i, x_j)$ is given by the \textit{radial basis function} (rbf) kernel $K_\mathrm{rbf}(x_i, x_j) \equiv \exp(-\gamma |x_i - x_j|^2) \, , \gamma > 0$. By definition, this kernel can be interpreted as a rescaled similarity measure of two samples $x_i$ and $x_j$.

After finding the support vectors $\alpha_i, i=1, \dots, l$, the decision function $f(x)$ is obtained by

\begin{equation}\label{eq:ocsvm_decision}
    f(x) = \mathrm{sgn}\left( \sum\limits_{i=1}^l \, \alpha_i K(x_i, x) - \varrho \right) \, .
\end{equation}
Details on computational aspects of solving the quadratic optimization problem in Eq.\eqref{eq:ocsvm_dual} can be found in references \cite{chang2011libsvm,steinwart}. \\

The precision $P$ and recall $R$ are useful measures to evaluate the success of model predictions when the classes are very imbalanced. Those metrics are defined by the following ratios

\begin{align}
    P &= \frac{\mathrm{TP}}{\mathrm{TP} + \mathrm{FP}} \, , \\
    R &= \frac{\mathrm{TP}}{\mathrm{TP} + \mathrm{FN}} \, ,
\end{align}
where the notation indicates true positive (TP), false positive (FP) and false negative (FN) predictions. Depending on the application either precision or recall can be of higher importance. However, here we treat them equally important and hence evaluate the overall performance of the model with the F1 score, being the harmonic mean of the precision $P$ and the recall $R$
\begin{equation}
    \mathrm{F1} = \frac{2}{P^{-1} + R^{-1}} \, . \label{eq:f1_score}
\end{equation}

\subsection*{B. Projected Quantum Kernels}

Embedding data into a quantum Hilbert space enables access to high-dimensional feature maps, where recent works have shown that specific learning tasks can be solved with an advantage over classical models \cite{liu2021, kernel_glick2024covariant, kernel_bowles2024better}. However, introducing too many dimensions, however can lead to the curse of dimensionality, introduced in Sec. I. In such a case the model becomes too expressive and kernel matrix elements concentrate exponentially towards the same value \cite{QSVMconcentration}, leading to overfitting. To control the dimensionality of the quantum feature space and the potential of overfitting, it has been shown that a projection of the embedded data back into the classical feature space, as proposed in \cite{huang}, can be sufficient. One way to achieve this is through the use of one-qubit reduced density matrices (1Q-RDM) $\rho_k$, where instead of the full density matrix all but the $k$th qubit are traced out. The 1Q-RDM can be recombined into new kernels, called projected quantum kernels. For example the projected quantum radial basis function (qrbf) kernel is defined via
\begin{equation}\label{eq:qrbf}
    K_\gamma\left(x_i, x_j\right) \equiv \exp\left( -\gamma \sum\limits_{k=1}^N \, \left|\left| \rho_k\left(x_i\right) - \rho_k\left(x_j\right) \right|\right|_\mathrm{F}^2 \right) \, ,
\end{equation}
where $||.||_\mathrm{F}$ is the Frobenius norm, $\gamma > 0$ and $N$ the number of qubits corresponding to the feature map. This alone does not avoid the concentration phenomenon in Ref.~\cite{QSVMconcentration}. However, as we we do not optimize the feature maps corresponding to our kernels, we do not need to circumvent that phenomenon.

\subsection*{C. Partial State Tomography}

The projected quantum kernels, as in Eq.~\eqref{eq:qrbf}, require knowledge about the 1Q-RDMs $\rho_k(x)$ of the state $\rho(x)$. Those can be obtained from real quantum hardware, via (partial) state tomography. This protocol is realized by performing the Pauli random measurement basis transformation $U \in \left\{ H, H S^\dagger, \mathds{1} \right\}$ on each qubit, as described in Ref. \cite{huang_shadows}, where the Hadamard gate $H$ and $S^{\dagger} = \mathrm{diag}(1, -i)$. Those unitaries are used to transform the default $Z$-basis measurements into $X$- and $Y$-basis measurements. For implementational reasons (on the IBM Quantum, IonQ and AQT backends we used), a fixed set of circuits has to be sent. Therefore, measurements on each qubit are performed where the shots are distributed equally on each Pauli basis, respectively. The measurement outcomes $b^{(1)}_k, b^{(2)}_k, \dots b^{(n)}_k$ are collected for each measurement basis on the $k$-th qubit $U_k$ and those tuples $(b^{(i)}_k, U_k)$ are mapped to snapshots $\hat{\rho}^{(i)}_k$ of each 1Q-RDM with
\begin{equation}\label{eq:tomography}
    \hat{\rho}^{(i)}_k = 3 \, U_k^\dagger \ket{b^{(i)}_k} \bra{b^{(i)}_k} U_k - \mathds{1} \, .
\end{equation}

Eventually, after averaging over all snapshots $S(k,n) = \left\{ \hat{\rho}^{(1)}_k, \hat{\rho}^{(2)}_k, \dots, \hat{\rho}^{(n)}_k \right\}$ an approximation of the 1Q-RDM of the $k$-th qubit is obtained.\\

\noindent Let us highlight that, in general, having access to $\rho_k$ is very useful, since it enables us to construct several other quantum kernels a posteriori. For example, the hyperparameter tuning for the kernel in Eq.~\eqref{eq:qrbf} doesn't require repeating the tomography.

\section*{III. RESULTS}

\subsection*{A. Benchmark Simulations}

\begin{figure*}[h!]
    \centering
    \includegraphics[width=\linewidth]{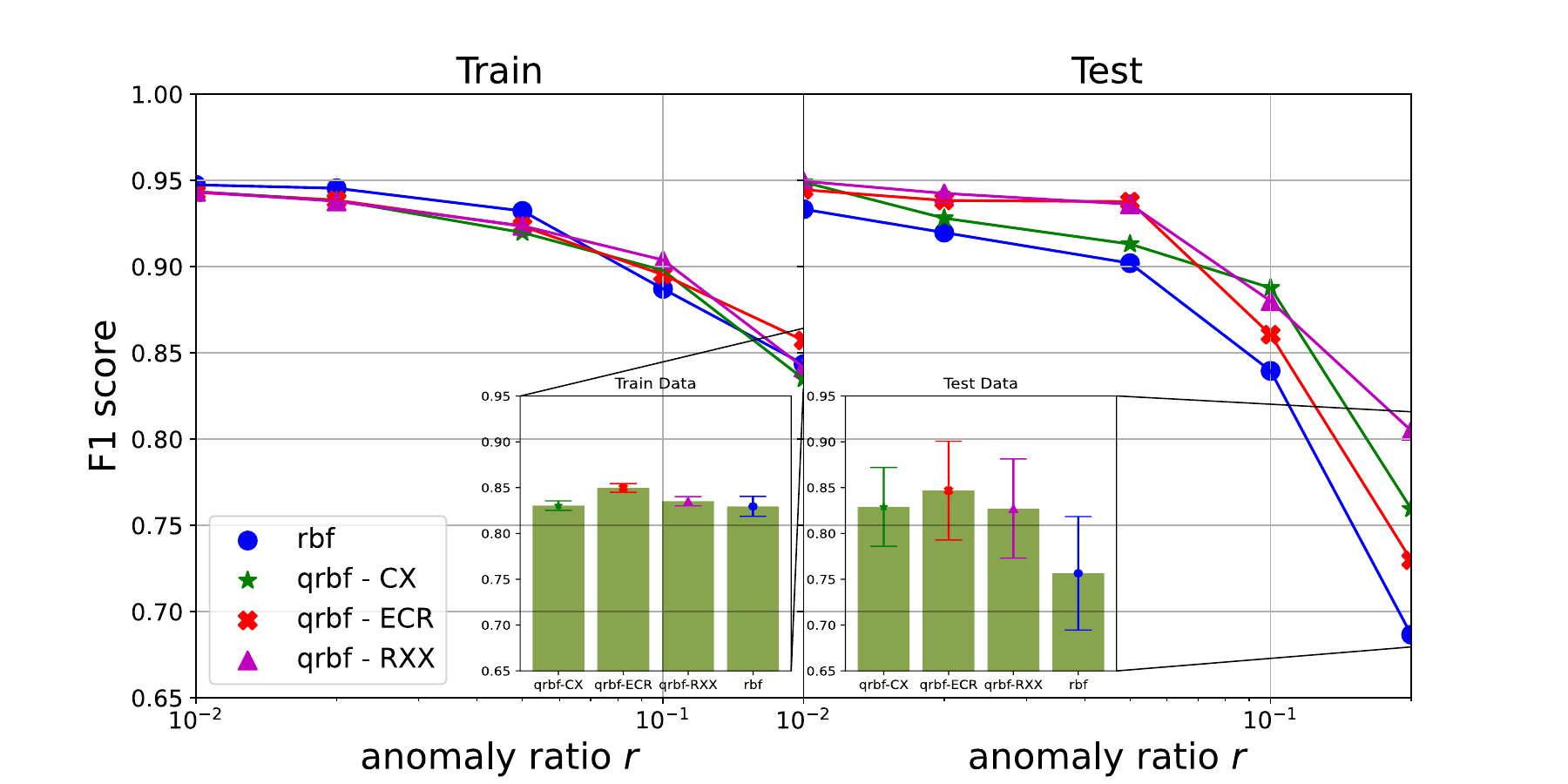}
    \caption{F1 score of the (q)rbf-OCSVM models in dependence of the anomaly ratio $r$ for the train (left) and test (right) split of the subsampled Sparkov dataset (test:train = 1:9). The qrbf-OCSVM consistenly outperforms the rbf-OCSVM for all $r$ on the test data for this split. The inset shows the mean F1 score and its standard deviation from a 10-fold cross-validation of the dataset for $r=0.2$. The estimated probabilities of the qrbf-OCSVM models surpassing the rbf-OCSVM model can be found in Tab.~\ref{tab:A_stats}.}
    \label{fig:benchmark_simulation}
\end{figure*}

For our simulations, we subsampled the Sparkov dataset of credit card transactions ~\cite{Leroy.2020} as follows (for more details cf. App. B): We keep 100 anomalous samples and add normal data successively, such that the anomaly ratio $r$ (ratio of anomalous samples with respect to the total number of data points) can be varied in a controlled way. Each sample $x_i$ is encoded into a quantum circuit for each of the three hardware-efficient feature maps. More details concerning the concrete feature maps can be found in App. A. In our analysis, we simulated the partial tomography protocol, explained in Sec. II.B for each sample with equally distributed Pauli measurements, which can be implemented on a gate-based quantum computer by evolving each qubit with one of the single-qubit unitaries $H, H S^\dagger, \mathds{1}$. Together with the measurement outcomes $b^{(1)}_k, \dots, b^{(200)}_k$ the snapshots $\hat{\rho}^{(1)}_k, \dots, \hat{\rho}^{(200)}_k$ are obtained with Eq.~\eqref{eq:tomography}. In detail, we obtained an approximation of the 1Q-RDMs by averaging over all snapshots $\{ \hat{\rho}^{(1)}_k, \dots, \hat{\rho}^{(200)}_k \}$ for each qubit $k$. Using Eq.~\eqref{eq:qrbf}, we obtained the qrbf kernels and inserted them into the OCSVM problem (cf. Eq.~\eqref{eq:ocsvm_dual}). By solving the dual optimization problem in Eq.~\eqref{eq:ocsvm_dual} we obtain the support vectors $\alpha_1, \dots \alpha_l$ needed for the classifier Eq.~\eqref{eq:ocsvm_decision}. For reference, we compare the results to the standard classical rbf-OCSVM. The results are shown in Fig.~\ref{fig:benchmark_simulation}. For comparison, a model that declares every sample as normal achieves a F1 score of $2r/(r+1)$, which equals $1/3$ for $r=0.2$. All (q)rbf-OCSVM in use have the parameters $\nu = 0.1$ and $\gamma = 0.1$ for every $r$ Those values for $\nu$ and $\gamma$ have been the best combination from a grid search in a 5-fold cross validation with a grid consisting of $\gamma \in \{ 10^{-4}, 10^{-3}, 10^{-2}, 10^{-1}, 10^{0}, 10^{1} \}$ and $\nu \in \{ 10^{-2}, 10^{-1}, 10^{0} \}$ for $r=10^{-2}$ for all feature maps. The used feature maps are shown in the App. A in Fig.~\ref{fig:he_fms}. For $r < 5 \cdot 10^{-2}$ one identifies a F1 score above $0.9$. In practice, such a score is considered a performant anomaly detection model. Further decreasing $r$ the F1 score saturates towards $\mathrm{F1} \approx 0.95$.\footnote{\noindent We ran the rbf-OCSVM also for the full dataset (8,291 anomalies, 1,791,345 normal samples) with $r=0.004607$ and obtained $\mathrm{F1} \approx 0.95$ for both the train and test data (test:train = 1:9).}. In our experiments, we observe that the qrbf-OCSVMs are very likely outperforming the rbf-OCSVM for all $r$ on the test data. This gap between both models is even bigger for higher $r$, where the model is trained on significantly less data ($r=0.2$ amounts to 500 samples whereas $10^4$ samples for $r=0.01$) as it was already hinted in \cite{caro2022generalization} for QML models in general. To generate more insight into the likelihood of the qrbf-OCSVMs outperforming the rbf-OCSVM model, we repeated the experiment among 10 different splits of the dataset with the same train-test ratio 9:1 for $r=0.2$ where the gap is most visible. By treating the F1 score of the classifier $C_\mathrm{F1}$ in this setting as a random variable under the truncated normal distribution with mean $\mu(C_\mathrm{F1})$ and standard deviation $\sigma(C_\mathrm{F1})$ on the interval $[0,1]$, we can calculate the probability of the qrbf-OCSVM classifier outperforming the rbf-OCSVM classifier by
\begin{equation}
    \mathrm{Pr}\left[ C^\mathrm{qrbf}_\mathrm{F1} - C^\mathrm{rbf}_\mathrm{F1} > 0 \right] =  \frac{1}{\mathcal{N}} \int\limits_0^1 \, \exp\left[-\frac{\left(x - \tilde{\mu}\right)^2}{2\tilde{\sigma}^2}\right] \mathrm{d}x \, , \label{eq:p_quantum_classical_gap}  
\end{equation}
with $\tilde{\sigma} \equiv \sigma\left(C^\mathrm{qrbf}_\mathrm{F1} - C^\mathrm{rbf}_\mathrm{F1}\right) = \sqrt{\sigma^2\left(C^\mathrm{qrbf}_\mathrm{F1}\right) + \sigma^2\left( C^\mathrm{rbf}_\mathrm{F1} \right)}$, $\tilde{\mu} \equiv \mu\left( C^\mathrm{qrbf}_\mathrm{F1} - C^\mathrm{rbf}_\mathrm{F1} \right) = \mu\left(C^\mathrm{qrbf}_\mathrm{F1}\right) - \mu\left(C^\mathrm{rbf}_\mathrm{F1}\right)$ and the normalization constant $\mathcal{N} \equiv 1/\sqrt{2 \pi \tilde{\sigma}^2} \left[ \Phi\left(\frac{1-\tilde{\mu}}{\sigma}\right) - \Phi\left( -\frac{\tilde{\mu}}{\tilde{\sigma}} \right) \right]$, where
\begin{equation}
    \Phi(x) \equiv \frac{1}{2} \left( 1 + \mathrm{erf}\left(\frac{x}{\sqrt{2}} \right)\right) \, , \label{eq:cdf}
\end{equation}
is the cumulative distribution function and $\mathrm{erf}(x) \equiv 2/\sqrt{\pi} \int\limits_0^{x} \, \exp(-t^2) \, \mathrm{d}t$ the Gaussian error function. We observed that all the estimated probabilities are above 64 \%, cf. Tab.~\ref{tab:A_stats}. 

\begin{table}
	\centering
	\caption{Probability that the qrbf-OCSVM model achieves a better F1 score than the rbf-OCSVM on the test data. The mean and variance of the $C^\mathrm{(q)rbf}_\mathrm{F1}$ were obtained for $r=0.2$.}
	\begin{tabular}{|c|c|c|c|}\hline
	Feature Map & $\mu\left( C_\mathrm{F1} \right)$ & $\sigma\left( C_\mathrm{F1} \right)$ & $\mathrm{Pr}\left[ C^\mathrm{qrbf}_\mathrm{F1} - C^\mathrm{rbf}_\mathrm{F1} > 0 \right]$  \\ \hline\hline
	rbf & 0.757 & 0.062 & none \\ \hline
	qrbf-CX & 0.829 & 0.043 & 69.2 \% \\ \hline
	qrbf-ECR & 0.847 & 0.059 & 74.7 \% \\ \hline
	qrbf-RXX & 0.827 & 0.054 & 64.8 \% \\ \hline 
	\end{tabular}
	\label{tab:A_stats}
\end{table}

\subsection*{B. State Tomography on Real Quantum Hardware}

\begin{figure*}[h!]
    \centering
    \includegraphics[width=\linewidth]{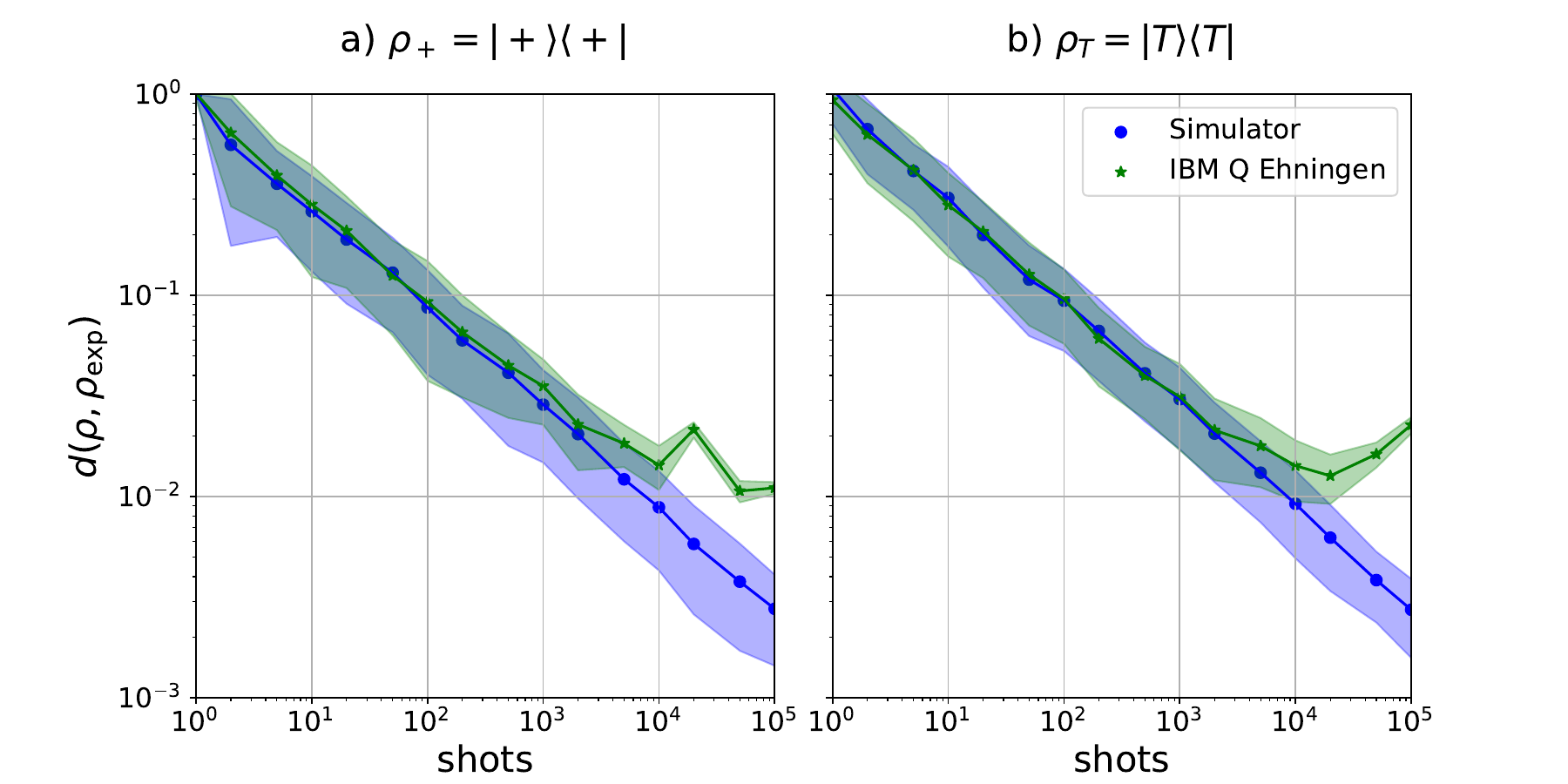}
    \caption{State tomography results for a) $\rho_{+}$ and b) $\rho_{T}$ (right) on a simulator with finite-sampling noise (blue) and on a Falcon r2 QPU by IBM Quantum (green). The experiments were repeated 100 times to estimate the standard deviation of the tomography precision. The operator 2-norm differences (cf. Eq.~\eqref{eq:op2norm}) between the true $\rho$ and $\rho_{\mathrm{exp}}$ coincide until shots on the order of $\approx 10^3$. While on the simulator $d(\rho, \rho_{\mathrm{exp}})$ continues to decrease polynomially, on the QPU a limit is reached due to the drift of the QPU calibration that becomes more evident for longer experiment runtimes.}
    \label{fig:tomography}
\end{figure*}

The quality of the results on real quantum hardware depends strongly on the quality of the partial shadow tomography. We use the same tomography protocol as in section III.A. In what follows, we quantify the precision of the partial state tomography by the difference between the true density matrix $\rho$ and the one obtained from the tomography experiments $\rho_\mathrm{exp}$ in the Frobenius norm

\begin{equation}\label{eq:op2norm}
    d(\rho, \rho_\mathrm{exp}) \equiv \sqrt{ \mathrm{Tr}\left[ (\rho - \rho_\mathrm{exp})^\dagger (\rho - \rho_\mathrm{exp}) \right] } \, .
\end{equation}

To test the performance of the state tomography protocol on the real quantum hardware, we prepared $\ket{+} = (\ket{0} + \ket{1})/\sqrt{2}$ and $\ket{T} = (\ket{0} + \exp(i\pi/4) \ket{1})/\sqrt{2}$ on a Falcon r2 QPU with 27 qubits by IBM Quantum. The result is shown in Fig.~\ref{fig:tomography}. For each number of shots, we repeat the experiment 100 times to estimate the variance of the tomography precision. Until shot numbers $\approx 10^3$ the simulation and the experiment on the QPU deliver results of roughly equal precision. However, further increasing the shot number only increases the precision on the simulator, whereas on the real hardware the precision saturates towards a limit. We observe that repeating the experiment on the same QPU at different times can lead to different results due to a drift of the calibration on the machine. This makes it challenging to reproduce the results, especially when the queues are long and waiting times can reach many days. While for a smaller number of shots the experiments (end-to-end) ran in a matter of seconds it took ~2.5 hours for $10^{5}$ shots. As long as the job is executed on the hardware, there is no recalibration in-between. Hence, there can be a drift of the system's properties and a worse state preparation and measurement (SPAM) during the run. \\
In practice, this means that besides finite coherence times and other crucial system properties, that influence successful experiment executions, it is important to shorten the total duration of the experiments on a current QPU, as well.

\subsection*{C. Performance on Real Quantum Hardware}

\begin{figure*}
    \centering
    \includegraphics[width=\linewidth]{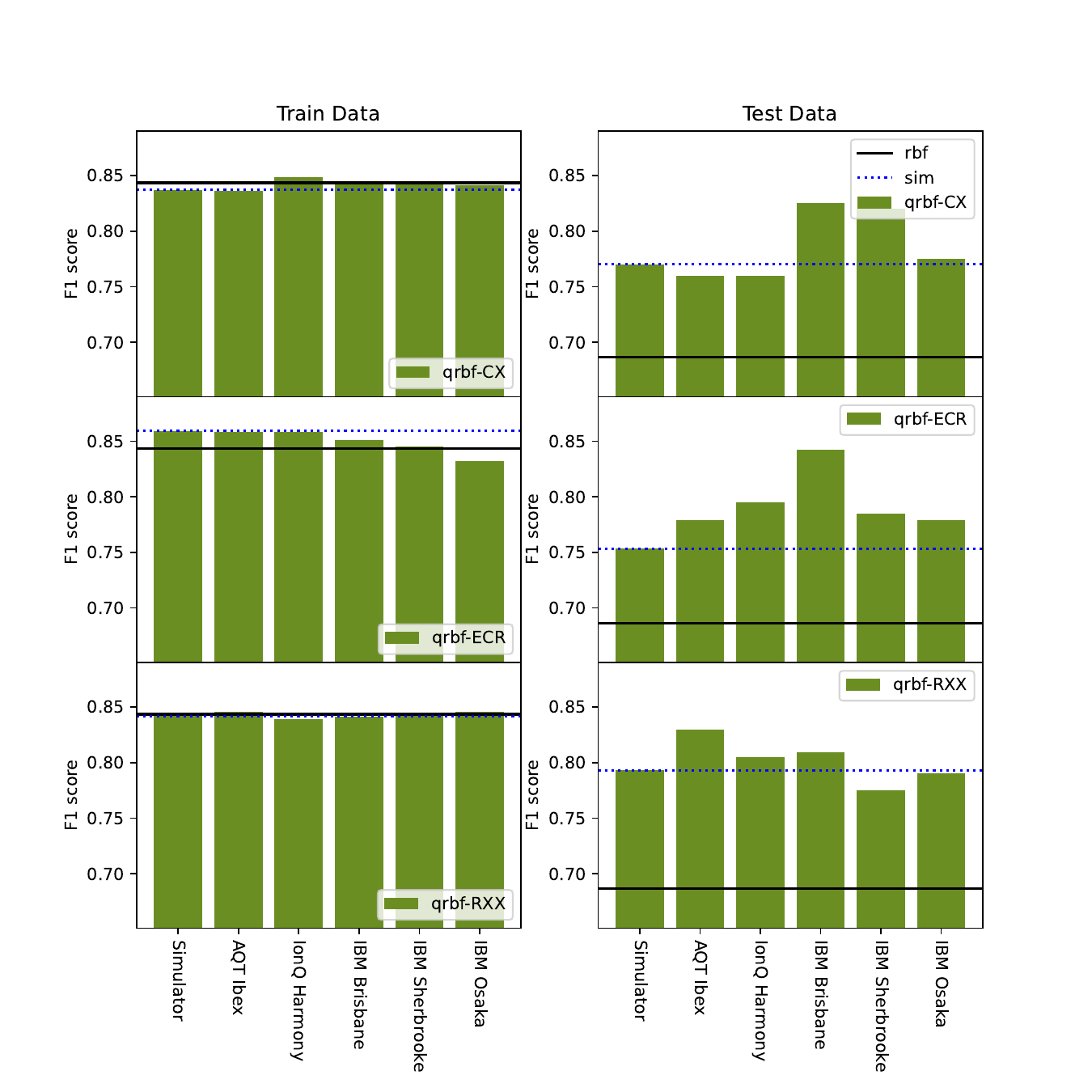}
    \caption{Hardware benchmark results for the (q)rbf-OCSVM models on a noise-free simulator, ion-trapped (AQT Ibex and IonQ Harmony) and superconducting (IBM Eagle r2/r3) quantum processors. The experiments were run on the same dataset as in Sec. III.A for $r=0.2$. The generalization enhancement by the qrbf models persists when transfering the algorithms to the real quantum hardware. The fluctuations of the model predictions result from the deviations between the true and experimental state $d(\rho, \rho_\mathrm{exp}) \neq 0$ obtained fom the tomography protocol.}
    \label{fig:hardware_results}
\end{figure*}

Finally, we report the experimental results of the anomaly detection task introduced in Sec. III.A on ion-trap and superconducting quantum computers. For the experiments, we had access to the backends AQT Ibex, IonQ Harmony (both ion-trap quantum computers) and IBM Eagle r2/r3 and Falcon r2 (superconducting quantum computers). To that end, we applied the qrbf-OCSVM model with the same three different hardware-efficient feature maps and number of shots as in the simulations in Sec. III.A for anomaly ratio $r=0.2$. The performance was evaluated via the F1 metric, defined in Eq.~\eqref{eq:f1_score}, on the train and the test split of the data (500 samples from the feature-engineered Sparkov dataset split into 450 train and 50 test samples), cf. App. B. We extract the 1Q-RDMs from the quantum computers by the same tomography protocol, as described in Sec. II.C and III.B. The resulting 1Q-RDMs are used for the calculation of the projected quantum kernels for the three feature maps \textit{qrbf-CX, qrbf-ECR} and \textit{qrbf-RXX}, using $\gamma = 0.1$. The results are depicted in Fig~\ref{fig:hardware_results}.

The F1 score of the qrbf-OCSVM fluctuates around the F1 score obtained by the noise-free simulator. This is due to the deviation resulting from the tomography $d(\rho,\rho_\mathrm{exp}) \neq 0$ that can lead to slightly different kernels. By chance, this results in either better or worse F1 scores. To shed more light into this, we repeated the simulation 128 times to estimate the standard deviation in the noise-free regime. This is the regime in which finite sampling is the only cause for fluctuations of the model predictions. Furthermore, we want to emphasize that these fluctuations are fundamentally different than those encountered in Sec. III.A., where we presented a more exhaustive model evaluation via 10-fold cross-validation. We report a neglible standard deviation ($\sigma(C_\mathrm{F1}) < 3 \cdot 10^{-3}$) for all the feature maps, cf. Table~\ref{tab:hardware_results_var}. Hence, the major cause of the fluctuation is not finite sampling but the hardware noise. Due to the use of shallow and hardware-efficient feature maps, we observed a generalization enhancement on the QPUs despite the noise. In fact, the predictions of the models that were executed on the QPUs is on par with those from the simulation for the train data and sometime even better on the test data. However, this is just one experiment execution of the whole QML models on the QPU and hence we can't properly evaluate the models. Therefore, further experimental data analysis is needed to confirm a generalization enhancement of quantum kernel methods in the presence of hardware noise. 

\begin{table}[]
    \centering
    \caption{Standard deviation of the F1 score for the qrbf-OCSVM models.}
    \begin{tabular}{|l|c|c|} \hline
        Feature Map & $\sigma(C_\mathrm{F1})$ Train & $\sigma(C_\mathrm{F1})$ Test \\ \hline\hline
        qrbf-CX & 0.00209 & 0.01457 \\ \hline
        qrbf-ECR & 0.00298 & 0.01989 \\ \hline
        qrbf-RXX &  0.00274 & 0.01320\\ \hline
    \end{tabular}
    \label{tab:hardware_results_var}
\end{table}

\section*{IV. Summary \& Outlook}

We conclude our work by summarizing our findings in the context of the research questions, which have been defined in Sec.~I. 

The simulations of our (q)rbf-OCSVM models on our 20-dimensional financial dataset have shown evidence that with a high probability the quantum rbf- models generalized significantly better than the classical counterparts in the small data and high anomaly regime. Further research is required to assess their performance across diverse data regimes, including lower and higher dimensions, as well as varying train-test set sizes. We have further demonstrated that this generalization enhancement is robust to hardware noise and finite sampling effects. However, due to limited access to real quantum hardware, a rigorous statistical analysis of these results remains an ongoing research area. As outlined in research questions 3 and 4, we systematically analyzed both encoding schemes and tomography precision to optimize our quantum model designs. To ensure compatibility with the quantum hardware, our data encodings were designed to minimize circuit depth. Our state tomography protocols employed Pauli random measurements, which, relying solely on single-qubit operations, were comparable in precision and sample complexity to the simulated tomography protocol considering only finite sampling effects. Experimental results revealed that, beyond device properties, experiment duration also significantly impacts tomography precision. Indeed, we observed a saturation and subsequent degradation of precision with increasing experiment time, likely attributable to device calibration drift. While the Pauli basis is well-suited for current hardware, exploring alternative bases could potentially yield more efficient protocols in terms of sample complexity. Further investigation into the performance of state tomography with informationally-complete POVMs is essential. While physical POVM realizations using projection-valued measurements and ancilla qubits are possible, their cost and feasibility require further study.

\section*{ACKNOWLEDGEMENTS}

This article was written as part of the Qu-Gov project, which was commissioned by the German Federal Ministry of Finance. The authors want to express their gratitude to Kim Nguyen, Manfred Paeschke and Oliver Muth for their continuous encouragement and support. Additional acknowledgements go to Christine Maier, Florian Girtler and David Vettese at Alpine Quantum Technologies for their great support.

\section*{BIBLIOGRAPHY}

\printbibliography[heading=none]

\section*{Appendix}

\subsection*{A. Hardware Efficient Feature Maps}

We use quantum hardware by AQT, IonQ and IBM Quantum. While IBM Quantum operates on a superconducting fixed-frequency transmon qubit platform, IonQ and AQT use ion traps. Each platform has its own unique set $\mathcal{N}$ of native gates which are realized within the physical system. These gates are optimized on a low level and each quantum circuit gets compiled to this set. \\

The chip connectivity poses a constraint on the types of feature maps that can be executed without too much overhead after compilation. We note, that trapped-ion platforms have an all-to-all connectivity while superconducting platforms only realize two-qubit gates on nearest neighbours. In order to still get comparable results between models that were obtained from either platform, we restrict our experiments to only feature maps that make only use of single-qubit gates and two-qubit gates on nearest-neighbours. 

\begin{figure}
    \centering
    \begin{subfigure}{0.45\textwidth}
        \includegraphics{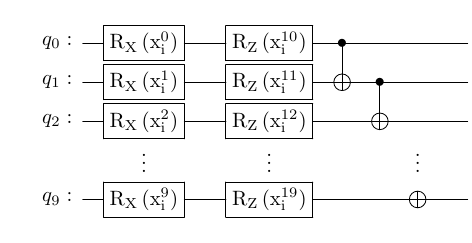}
        \caption{Feature Map: qrbf - CX}\label{fig:rx_rz_cx}        
    \end{subfigure}
    \centering
    \begin{subfigure}{0.45\textwidth}
        \includegraphics[width=\textwidth]{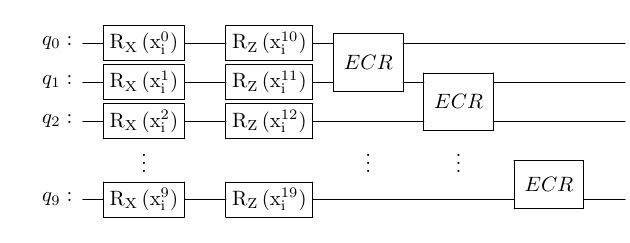}
        \caption{Feature Map: qrbf - ECR}\label{fig:rx_rz_ecr}
    \end{subfigure}
    \centering
    \begin{subfigure}{0.45\textwidth}
        \includegraphics[width=\textwidth]{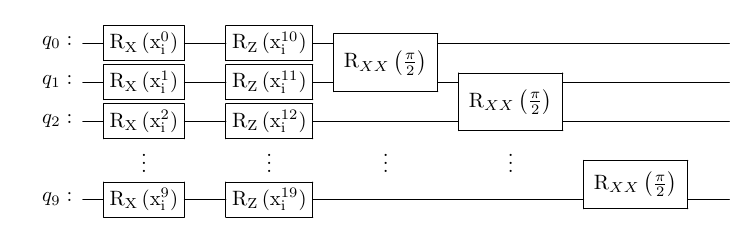}
        \caption{Feature Map: qrbf - RXX}\label{fig:rx_rz_rxx}
    \end{subfigure}
    \caption{These are the hardware efficient feature maps that were chosen for our experiments.}\label{fig:he_fms}
\end{figure}

Up until early 2024, the native gate set for IBM's device was mainly given by
\begin{equation}
    \mathcal{N} = \left\{ X, \sqrt{X}, R_Z(\lambda), CX  \right\} \, , \label{eq:native_cx}
\end{equation}
where $R_Z(\lambda) = \exp(-i \lambda Z/2)$. The corresponding hardware efficient feature map is shown in Fig.~\ref{fig:rx_rz_cx}, where $R_X$ is transpiled into $\sqrt{X}$'s and $R_Z$. 

\noindent Recent advances in variational quantum algorithms in the NISQ era motivate the preference of the pure echoed CR (ECR) over the CX gate since the ECR pulse is shorter than the CX pulse. Therefore, more recent IBM Quantum systems deploy an alternative native gate set 
\begin{equation}
    \mathcal{N} = \left\{ X, \sqrt{X}, R_Z(\lambda), ECR \right\}\, , \label{eq:native_ecr}
\end{equation}
where 
\begin{equation}
    \begin{split}
    ECR &= R_{ZX}\left( -\frac{\pi}{4} \right) \left( R_X(\pi) \otimes I \right) R_{ZX}\left( \frac{\pi}{4} \right)\, , \\
    R_X(\lambda) &= e^{-i \frac{\lambda}{2} X} \, , \\
    R_{ZX}(\lambda) &= e^{-i \frac{\lambda}{2} Z \otimes X} \, .
    \end{split}    
\end{equation}
The corresponding hardware efficient feature map is shown in Fig.~\ref{fig:rx_rz_ecr}, where $R_X$ is transpiled into $\sqrt{X}$'s and $R_Z$.\\

\noindent Ion trap based quantum computers however employ a different native get set as those on superconducting hardware. AQT uses 
\begin{equation}
    \mathcal{N} = \left\{ R_X(\alpha), R_Z(\beta), R_{XX}(\gamma) \right\} \, , \label{eq:native_rxx}
\end{equation}
where
\begin{equation}
    \begin{split}
        R_Z(\lambda) &= e^{-i\frac{\lambda}{2} Z}  \, , \\
        R_{XX}(\lambda) &= e^{-i \frac{\lambda}{2} X \otimes X} \, . 
    \end{split}
\end{equation}
The corresponding hardware efficient feature map is shown in Fig.~\ref{fig:rx_rz_rxx}. For $\lambda = \pi/2$ the $R_{XX}$-gate is maximally entangling. 

\noindent For the sake of realizing our models in the most efficient way possible, we design our feature maps, such that the hardware's respective native gate sets in Eq.~\eqref{eq:native_cx}, Eq.~\eqref{eq:native_ecr} and Eq.~\eqref{eq:native_rxx} are used.

\subsection*{B. Data and Feature Engineering}

\begin{figure}
    \centering
    \begin{subfigure}{0.45\textwidth}
        \includegraphics[width=\textwidth]{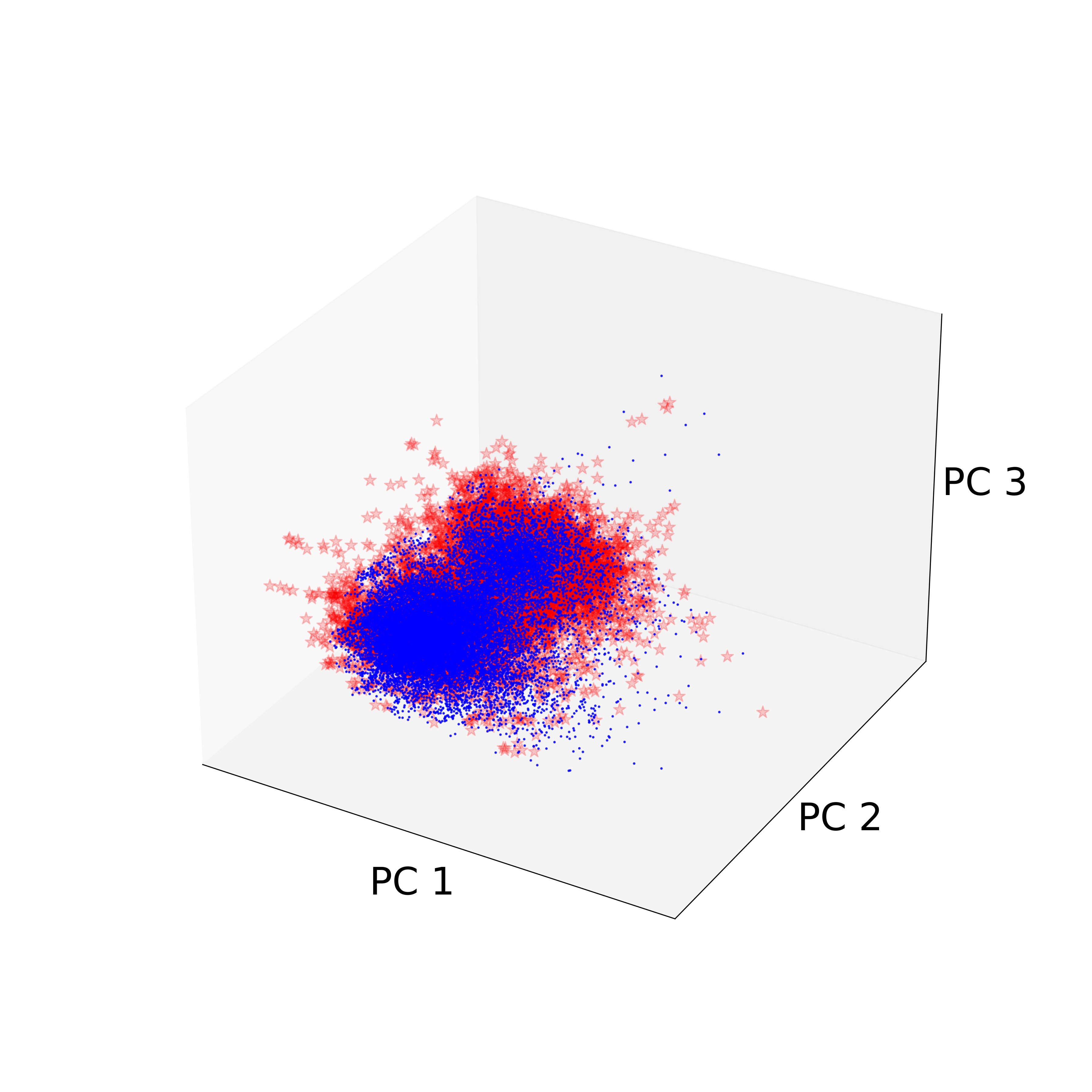}
        \caption{}
        \label{fig:pca_plot}
    \end{subfigure}
    \centering
    \begin{subfigure}{0.45\textwidth}
        \includegraphics[width=\textwidth]{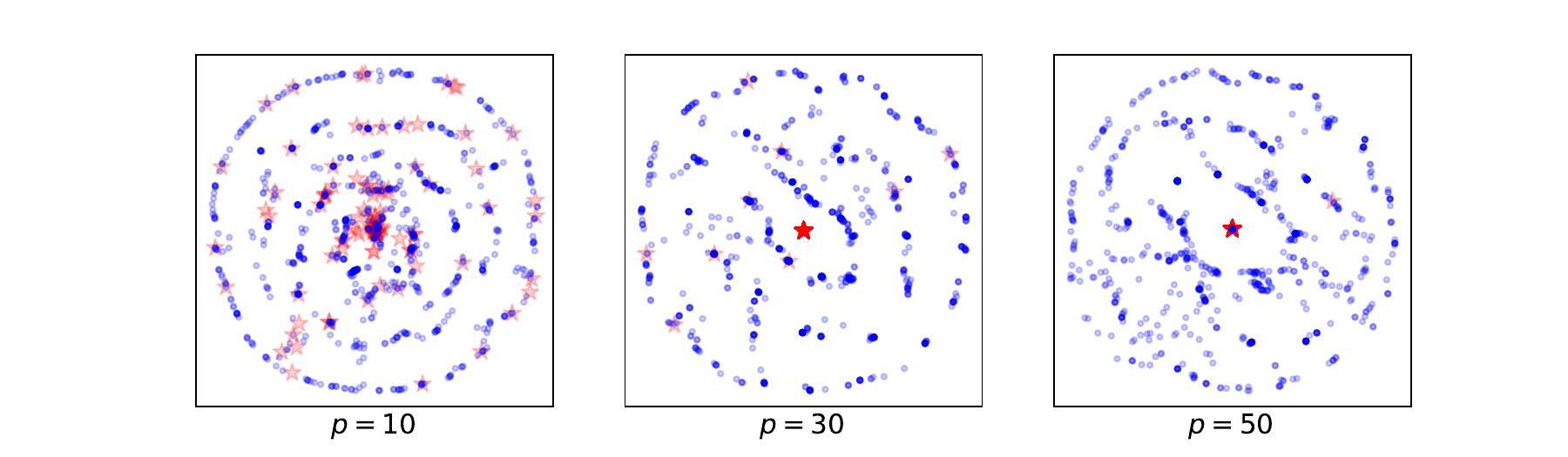}
        \caption{}
        \label{fig:tsne_plot}
    \end{subfigure}
    \caption{Visualizations of the feature engineered Sparkov dataset with a) all samples after principal component analysis and b) the first 2000 samples after t-SNE with perplexity $p \in \{10,30,50\}$. Red stars denote anomalies and blue dots normal samples. For better visibility we plot the red stars larger, since they make up only $r \approx 0.52 \%$ of the dataset.}\label{fig:visuals}
\end{figure}

The \textit{Sparkov} dataset \cite{Shenoy.2020} was synthetically generated by the \textit{Sparkov Data Generation} tool \cite{Leroy.2020} and contains data for two full years (2019 and 2020) for credit cards of 1,000 customers and 800 merchants. In total there are 1~852~394 samples with each 22 different features, including the label fraud or non-fraud. Neither missing data nor duplicates are present. The total data contains 9,651 fraud transactions ($\approx 0.52\%$), making it an unbalanced anomaly detection case. Using Cramér's V association coefficient \cite{Cramer.1946}, a slight association between the product category (14 cumulated categorial values such as »health and fitness« or »food and dining«) and the fraud state was observed (0.071) on the raw data and insignificant association values for all the other variables (0.037 at most).

By grouping the data by city of the card owner, inconsistencies were identified which did not occur for the zip codes of the card owners. This was interpreted as cities with equal names in different states (of which there are several in the USA), therefore it was continued using the zip codes for the following steps. 

\begin{figure}
    \centering
    \includegraphics[ width = \linewidth ]{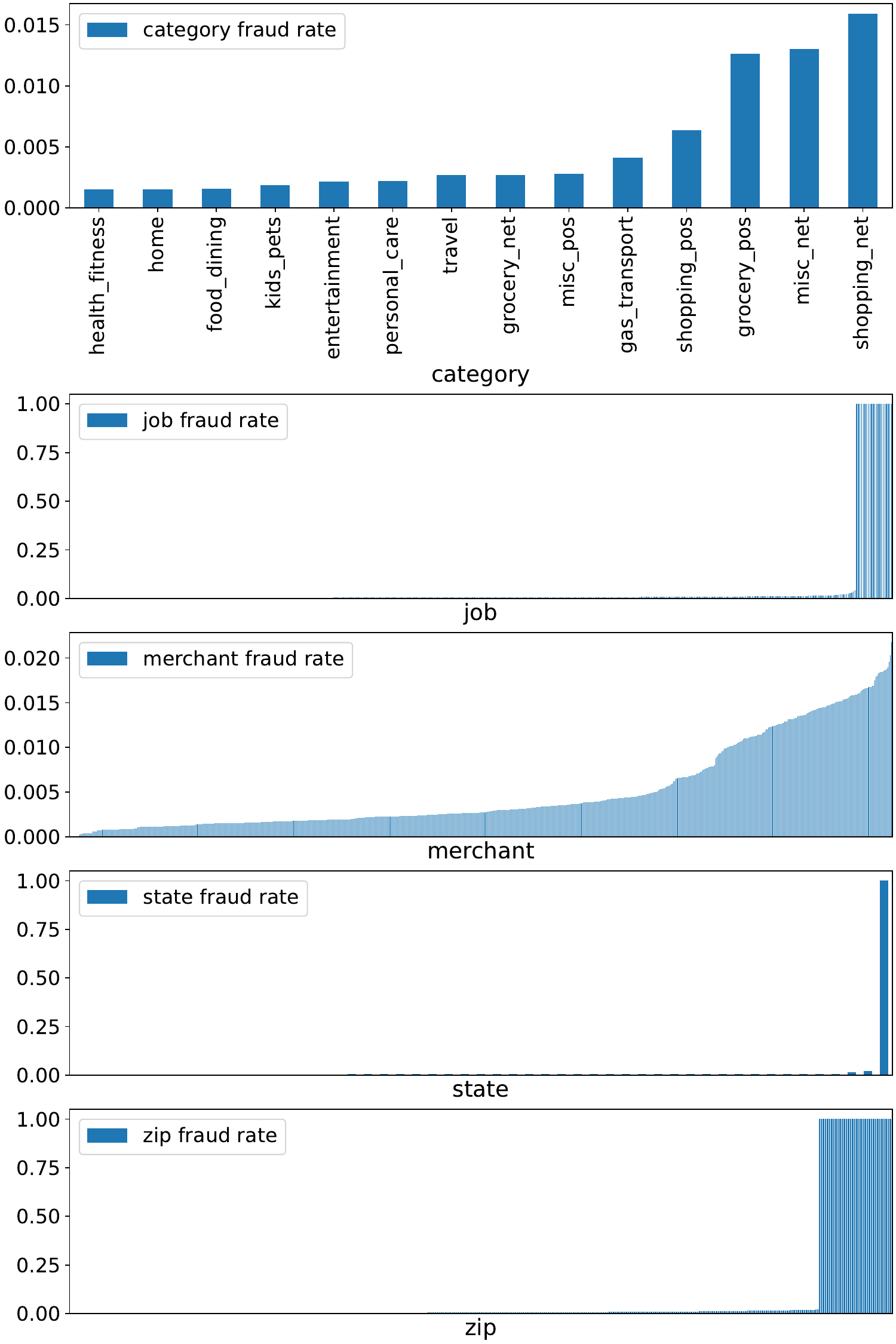}
    \caption{The fraud rates of the categorial variables in ascending order. It can be seen already here that the data at hand is simulated and not real, since in the real world the feature are not expected to behave like a step function. This artificialness in the data becomes most evident in the fraud rate vs job, state or zip code plots.}
    \label{fig:fraud_rates}
\end{figure}

Targeting the fraud transactions, the values were investigated one after the other as described in the following. Potentially relevant information content about the fraud state was observed by visual representation of the fraud rates for different variables: the zip code of the card owner, the merchant, the product category, and the job. The fraud rates of the categorial variables can be seen in \autoref{fig:fraud_rates}.

\begin{figure}
    \centering
    \includegraphics[ width = \linewidth ]{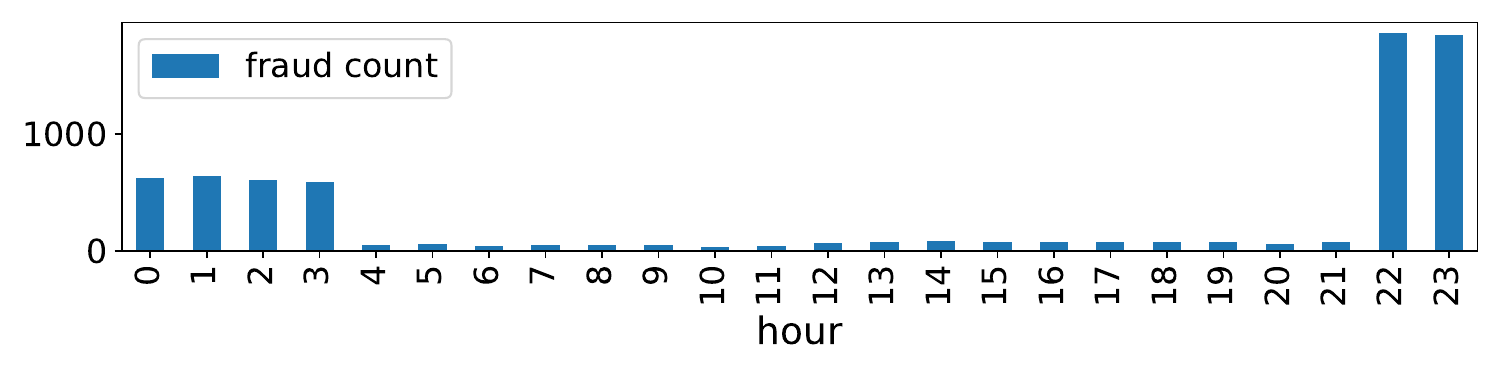}
    \caption{The fraud count grouped by hour of day. Clearly the data is biased towards frauds conducted during night.}
    \label{fig:frauds_per_time}
\end{figure}

A significant increase of fraud transactions was observed during night time (i.~e., between 9pm and 4am, cf. \autoref{fig:frauds_per_time}). The number of recent credit card transaction on the same card was also identified as a relevant indicator by grouping per credit card. Using the latitude and longitude of both, the transaction location and the merchant location seemed irrelevant, even if considering the distance between them. The low Pearson correlation coefficient (PCC) of the features and the label \textit{is\_fraud} can be seen in \autoref{tab:variable_table}. The differences were also investigated by the PCC but dropped, since no higher information value could be observed.

\begin{figure}
    \centering
    \includegraphics[ width = \linewidth ]{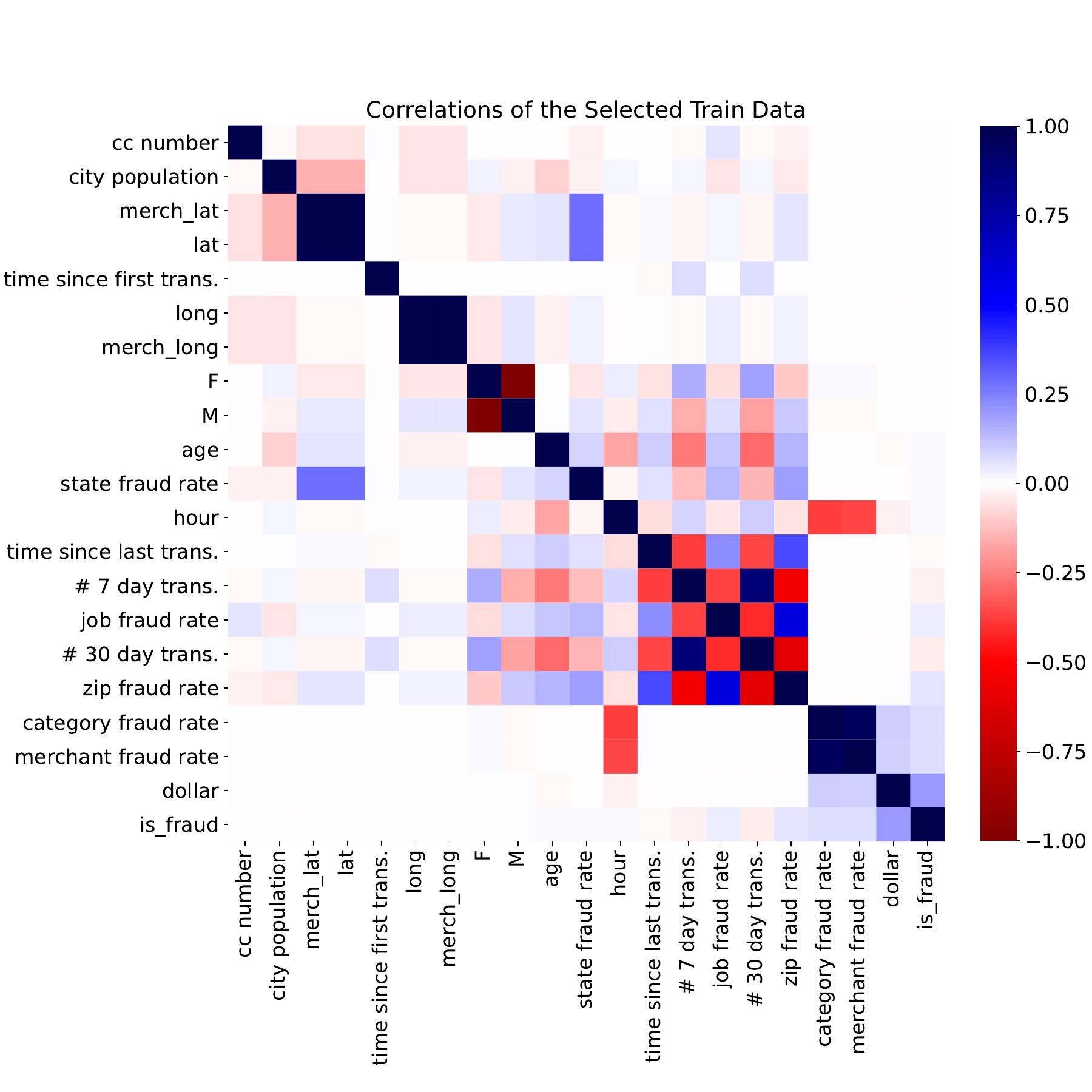}
    \caption{Heatmap of the PCC values shown in \autoref{tab:variable_table}.}
    \label{fig:pcc_heatmap}
\end{figure}

Due to the quantity of categories in several variables, fraud rates were used instead of dummy variables to prevent suffering from the curse of dimensionality. The fraud rates were calculated by the ratio of the number of fraud transactions divided by the number of samples after grouping by the respective features. The only categorial variable which was used with two binary dummies is the gender of the credit card holder. After dropping variables not longer used, the final dataset after the feature engineering consisted of 21 variables including the label. This results in the \textit{feature engineered Sparkov dataset} and is visualized in Fig.~\ref{fig:visuals}. The features are represented in \autoref{tab:variable_table} sorted by their Pearson correlation coefficient (PCC) \cite{Pearson.1895} with the label and can also be seen as a heatmap in \autoref{fig:pcc_heatmap}. \\


\begin{table}[!ht]
    \centering
    \begin{tabular}{p{5cm} c}
        \textbf{Variable Name}     & \textbf{PCC}\\\hline
        cc number                  & 0.000906 \\
        \rowcolor{fhggray}
        city population            & 0.000984 \\
        merch\_lat                 & 0.001136 \\
        \rowcolor{fhggray}
        lat                        & 0.001278 \\
        time since first trans.    & 0.001590 \\
        \rowcolor{fhggray}
        long                       & 0.001610 \\
        merch\_long                & 0.001627 \\
        \rowcolor{fhggray}
        F                          & 0.007707 \\
        M                          & 0.007707 \\
        \rowcolor{fhggray}
        age                        & 0.007883 \\
        state fraud rate           & 0.008171 \\
        \rowcolor{fhggray}
        hour                       & 0.011933 \\
        time since last trans.     & 0.015552 \\
        \rowcolor{fhggray}
        \# 7 day trans.            & 0.026886 \\
        job fraud rate             & 0.031661 \\
        \rowcolor{fhggray}
        \# 30 day trans.           & 0.033928 \\
        zip fraud rate             & 0.050097 \\
        \rowcolor{fhggray}
        category fraud rate        & 0.063912 \\
        merchant fraud rate        & 0.065614 \\
        \rowcolor{fhggray}
        dollar                     & 0.202042 \\
        is\_fraud                  & 1.000000
    \end{tabular}
    \caption{The variables after the feature engineering sorted by their PCC with the label. Abbreviations are as follows: »\#« denotes a count, »F« stands for female, »M« stands for male, »lat« stands for latitude, »long« stands for longitude.}
    \label{tab:variable_table}
\end{table}

\end{document}